*Review Article*

# The Study of Nebular Emission on Nearby Spiral Galaxies in the IFU Era

**Fernando Fabián Rosales-Ortega**

*Instituto Nacional de Astrofísica, Óptica y Electrónica, Luis E. Erro 1, 72840 Tonantzintla, PUE, Mexico*

Correspondence should be addressed to Fernando Fabián Rosales-Ortega; frosales@inaoep.mx

A new generation of wide-field emission-line surveys based on integral field units (IFU) is allowing us to obtain spatially resolved information of the gas-phase emission in nearby late-type galaxies, based on large samples of HII regions and full two-dimensional coverage. These observations are allowing us to discover and characterise abundance differentials between galactic substructures and new scaling relations with global physical properties. Here I review some highlights of our current studies employing this technique: (1) the case study of NGC 628, the largest galaxy ever sampled with an IFU; (2) a statistical approach to the abundance gradients of spiral galaxies, which indicates a *universal* radial gradient for oxygen abundance; and (3) the discovery of a new scaling relation of HII regions in spiral galaxies, the *local* mass-metallicity relation of star-forming galaxies. The observational properties and constrains found in local galaxies using this new technique will allow us to interpret the gas-phase abundance of analogue high-z systems.

## 1. Introduction

The study of the interstellar medium (ISM), like many other areas of astrophysics, has undergone a remarkable acceleration in the flow of data over the last few years. Large surveys such as the 2dFGRS [1], SDSS [2], GEMS [3], or COSMOS [4], to name a few, have revolutionised our understanding of the Universe and its constituents as they have enabled us to study the global properties of a large number of objects, allowing for meaningful statistical analysis to be performed, together with a broad coverage of galaxy subtypes and environmental conditions.

The nebular emission arising from extragalactic objects has played an important role in this new understanding. Nebular emission lines have been, historically, the main tool at our disposal for the direct measurement of the gas-phase abundance at discrete spatial positions in low redshift galaxies. They trace the young, massive star component in galaxies, illuminating and ionizing cubic kiloparsec-sized volumes of ISM. Metals are a fundamental parameter for cooling mechanisms in the intergalactic and interstellar medium, star-formation, stellar physics, and planet formation. Measuring the chemical abundance in individual galaxies and galactic substructures, over a wide range of redshifts, is a crucial step to understanding the chemical evolution and nucleosynthesis at different epochs, since the heavy atomic nuclei trace the evolution of past and current stellar generations. This evolution is dictated by a complex array of parameters, including the local initial gas composition, star-formation history (SFH), gas infall and outflows, radial transport and mixing of gas within discs, stellar yields, and the initial mass function. Although it is difficult to disentangle the effects of the various contributors, determinations of current elemental abundance constrain the possible evolutionary histories of the existing stars and galaxies, and the interaction of galaxies with the intergalactic medium. The details of such a complex mechanism are still observationally not well established and theoretically not well developed and threaten our understanding of galaxy evolution from the early Universe to the present day.

The relevance of the study of the ISM in the local Universe cannot be underestimated, since it actually constitutes the bases of the methods and calibrations employed to derive abundance and their relations with global galaxy parameters in high redshift galaxies (e.g., [5, 6]), objects that are typically solely identifiable by their emission line spectra. Nearby galaxies offer a unique opportunity to study the SFH-ISM coupling on a spatially resolved basis, over large dynamic ranges in gas density and pressure, metallicity, dust content, and other physically relevant parameters of gas and dust. However, most of the observations targeting nebular emission in nearby galaxies have been made with multibroad-band and narrow-band imaging in the optical and near-infrared, or single-aperture or long-slit spectrographs, resulting in samples of typically a dozen or fewer HII regions per galaxy. These observations have been used to derive the properties of their dominant stellar populations, gas content, and kinematics (e.g., [7–9]). Nevertheless, despite many efforts, it has been difficult to obtain a complete picture of the main properties of these galaxies, especially those ones that can only be revealed by spectroscopic studies (like the nature of the ionization and/or the metal content of the gas). This is because previous spectroscopic studies only sampled a very few discrete regions in these complex targets (e.g., [10, 11]), or used narrow-band imaging of specific fields to obtain information of star-forming regions and the ionized gas (e.g., [9]), and in many cases they were sampling very particular types of regions [12–15]. Integrated spectra over large apertures were required to derive these properties in a more complete way (e.g., *drift-scanning*, [16]), but even in these cases, only a single integrated spectrum is derived, and the spatial information is lost.

On the other hand, although large spectroscopic surveys like the 2dFGRS or the SDSS do provide a large number of objects sampled and vast statistical information, they are generally limited to one spectrum per galaxy, thus missing all the radial information and spatially resolved properties of the galaxy. These surveys have been successful to describe the integrated properties and relations of a large number of galaxies along a wide redshift range. But galaxies are complex systems not fully represented by a single spectrum or just broad band colours. Disc and spheroidal components are structurally and dynamically different entities with different SFH and chemical evolution. A main drawback of this technique is that it leads to aperture bias that is difficult to control, as the area covered to integrate the spectra corresponds to different physical scales at different redshifts (e.g., SDSS), and also the physical mechanisms involved in ionizing the gas may be very different within the sampled area, as this would include regions with emission due to diffuse ionized gas (DIG), shocks, or AGN/LINER activity.

The advent of Multi-Object Spectrometers (MOS) and Integral Field Spectroscopy (IFS) instruments with large fields of view (FoV) now offers us the opportunity to undertake a new generation of surveys, based on a full two-dimensional (2D) coverage of the optical extent of nearby galaxies. The first application of IFS to obtain spatially resolved, continuously sampled spectroscopy of certain portions of nearby galaxies was due to the SAURON project [17, 18]. SAURON was specifically designed to study the kinematics and stellar populations of a sample of nearby elliptical and lenticular galaxies. The application of SAURON to spiral galaxies was restricted to the study of spiral bulges [19]. However, IFS was rarely used in a "survey mode" to investigate sizeable samples. There were several reasons for the lack of a systematic study targeting galaxies in the local Universe using IFS that could cover a substantial fraction of their optical sizes. The reasons included small wavelength coverage, fibre-optic calibration problems, but mainly the limited FoV of the instruments available worldwide. Most IFUs have a FoV of the order of arcsec, preventing a good coverage of the target galaxies on the sky in a reasonable time, even with a mosaicking technique. Furthermore, in some cases the emission lines used in chemical abundance studies were not covered by the restricted wavelenght range of the instruments. Moreover, the complex data reduction and visualisation imposed a further obstacle.

In order to fill this gap, in the last few years we started a major observational programme aimed at studying the 2D properties of the ionized gas and HII regions in a representative sample of nearby face-on spiral galaxies using IFS. The spatially resolved information provided by these observations is allowing us to test and extend the previous body of results from small-sample studies, while at the same time it opens up a new frontier of studying the 2D gas abundance on discs and the intrinsic dispersion in metallicity, progressing from a one-dimensional study (radial abundance gradients) to a 2D understanding (distributions), allowing us at the same time to strengthen the diagnostic methods that are used to measure HII region abundance in galaxies.

Here we present the highlights of our current studies employing this large spectroscopic database: (1) the case of NGC 628, the largest galaxy ever sampled with IFS; (2) an IFS-based statistical approach to the abundance gradients of spiral galaxies; and (3) the discovery of a new scaling relation of HII regions in spiral galaxies and how we use it to to reproduce—with remarkable agreement—the mass-metallicity relation of star-forming galaxies.

## 2. A IFS Sample of Nearby Disc Galaxies

The studies here described were performed using IFS data of a sample of nearby disc galaxies. The observations were designed to obtain continuous coverage spectra of the whole surface of the galaxies. They include observations from the PPAK IFS Nearby Galaxies Survey: PINGS [20], and a sample of face-on spiral galaxies from Mármol-Queraltó et al. [21], as part of the feasibility studies for the CALIFA survey [22, 23], a legacy project which aims to observe a statistically complete sample of ~600 galaxies in the local Universe; all projects are carried out at the Centro Astronómico Hispano-Alemán of Calar Alto, Spain.

PINGS represented the first attempt to obtain continuous coverage spectra of the whole surface of a representative sample of late-type galaxies in the nearby Universe. This first sample includes normal, lopsided, interacting and barred spirals with a good range of galactic properties and star-forming environments with available multiwavelength public data (e.g., see Figure 1). The second sample consists of visually classified face-on spirals from Mármol-Queraltó et al. [21]

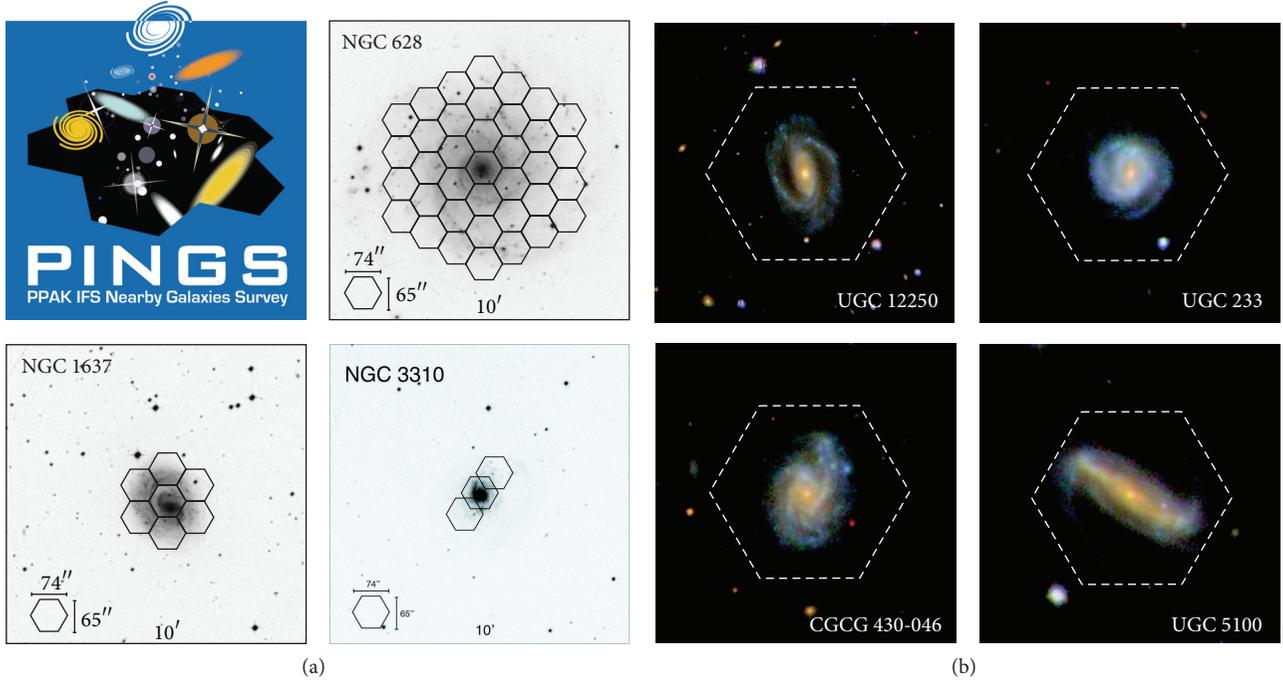

Figure 1: (a) Examples of the PINGS IFS mosaics, each panel shows a *B*-band Digital Sky Survey image of the galaxy with the PPAK mosaic pointings as overlaid hexagons indicating the FoV of the central fibre bundle. (b) Examples of the face-on spirals drawn from the Mármol-Queraltó et al. [21] sample of IFS galaxies, each panel shows a colour-composite SDSS image of the galaxy with the PPAK FoV footprint overlaid.

extracted from the SDSS DR4 imaging sample selecting galaxies brighter than r < 15.75 mag with redshifts in the range $0.005 < z < 0.025$ (selection in volume and limiting magnitude) and from face-on disc galaxies included in the DiskMass Survey [24] with appropriate sizes to fill the FoV of the PPAK instrument (angular isophotal-diameter selection, see below).

Both samples were observed with the PMAS spectrograph [25] in the PPAK mode [26, 27] on the 3.5 m telescope in Calar Alto with similar setup, resolutions, and integration times, covering their optical extension up to ∼2.4 effective radii within a wavelength range ∼3700–7000 Å. The PPAK fiber bundle consists of 382 fibers of 2.7 arcsec diameter each. Of these 382 fibers, 331 (the science fibers) are concentrated in a single hexagonal bundle covering a field-of-view of 74 × 64 $arcsec^2$, with a filling factor of ∼60%. The sky background is sampled by 36 additional fibers, distributed in 6 bundles of 6 fibers each, along a circle ∼72 arcsec from the center of the instrument FoV.

In the case of PINGS, the observations consisted of IFU spectroscopic mosaics for 17 spiral galaxies within a maximum distance of 100 Mpc; the average distance of the sample is 28 Mpc (for $H_0 = 73\,km\,s^{-1}\,Mpc^{-1}$). Most of the objects in PINGS could not be covered in a single pointing with IFS instruments, so a new observing-reduction technique had to be developed to perform accurate mosaicking of the targets. The spectroscopic mosaicking was acquired during a period of three years and the final data set comprises more than 50 000 individual spectra, covering in total an observed area of nearly 80 $arcmin^2$, and an observed surface without precedents by a IFS study up to that point (the case study of NGC 628 presented in Section 3 is based in the data of this survey). For the second sample, the galaxies were observed over fifteen nights in several observing runs. The main difference is that, for the latter sample, a single pointing strategy using a dithering scheme was applied, while, for the largest galaxies of the PINGS survey, a mosaic comprising different pointings was required. This is due to the differences in projected size, considering the different redshift range of both samples: the PINGS galaxies correspond to $z \sim 0.001$–0.003, while, for the face-on spirals, it is $z \sim 0.01$–0.025. Therefore, in both survey samples, the data extent corresponds to about ∼2 effective radii for all galaxies (The effective radius is classically defined as the radius at which one half of the total light of the system is emitted). So the final sample comprises 38 objects, with a redshift range between ∼0.001 and 0.025. Although this sample is by no means a statistical subset of the galaxies in the local Universe, it is a representative sample of face-on, mostly quiescent, and spiral galaxies at the considered redshift range (see Figure 1).

Data reduction was performed using R3D [31], obtaining as an output a data cube for each galaxy, with a final spatial sampling between 1-2 arcsec/pixel, which translates to a linear physical size between a few hundreds of parsecs to ∼1 kpc. Using this database we catalogued more than ≈2500 HII regions with good spectroscopic quality in all 38 galaxies, representing one of the largest and more homogeneous 2D spectroscopic HII region surveys ever accomplished.

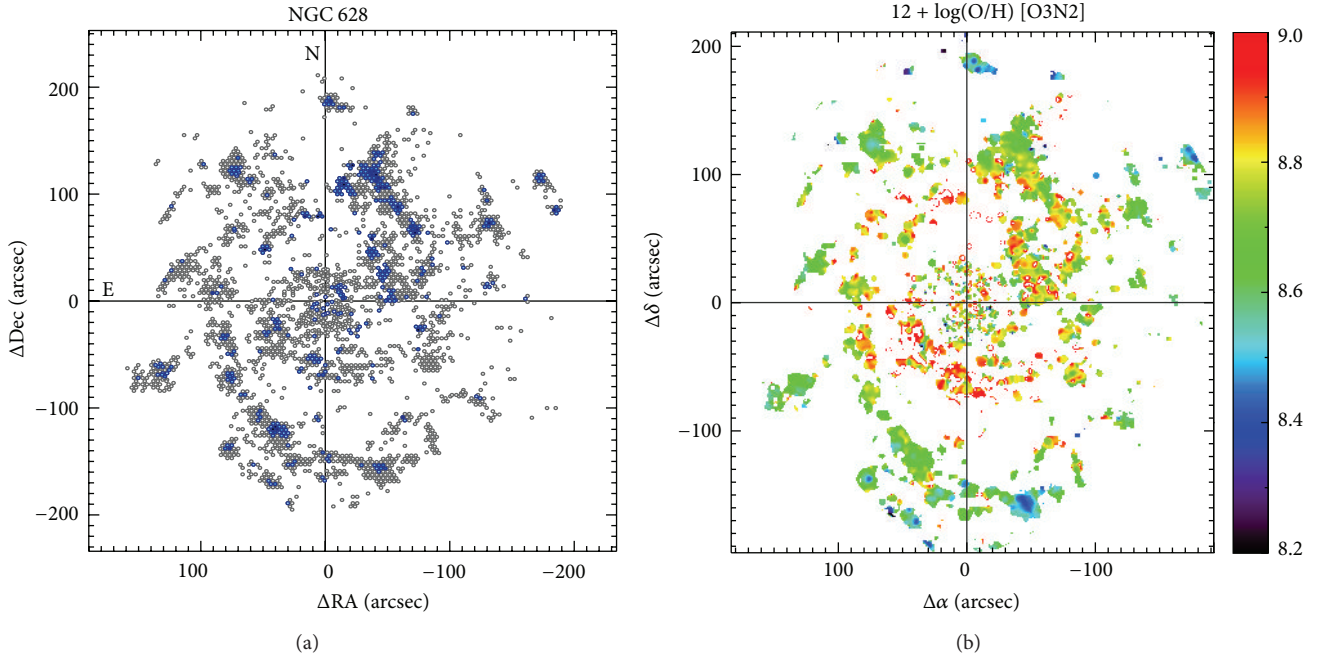

Figure 2: (a) Spatial map of the fibres within the IFS mosaic of NGC 628 where nebular emission was detected. Blue fibres indicate regions above a S/N threshold for a proper abundance analysis, and grey fibres correspond to a *diffuse* emission. The size and position of the fibres (at real scale) are displayed in the standard NE-positive orientation. The crosshairs mark the central reference point of the IFS mosaic. The colour intensity of each fibre in the blue sample has been scaled to the flux intensity of H$\alpha$ for that particular spectrum. (b) Oxygen abundance map of NGC 628 derived by applying the O3N2 calibrator [28] to the emission line maps of the galaxy. The figure shows a clear gradient in metallicity, with more abundant regions in the inner part or the galaxy. Figure adapted from Sánchez et al. [29] and Rosales-Ortega et al. [30].

The discussion presented in Sections 4 and 5 is based on these databases. The primary scientific objectives of these surveys were to use the 2D IFS observations to study the small and intermediate scale variation in the line emission and stellar continuum by means of pixel-resolved maps across the discs of nearby galaxies, as described in the following sections.

## 3. NGC 628: A Case Study of IFS-Based Nebular Emission Studies

NGC 628 (or M 74) is the largest galaxy in projected angular size (~10.5 × 9.5 arcmin$^2$, $z \sim 0.00219 \sim 9$ Mpc) of the PINGS sample. Due to the large size of NGC 628 compared to the FoV of the PPAK instrument (72 × 64 arcsec$^2$) a mosaicking scheme was adopted, employing 34 different pointings. The initial pointing was centered on the bulge of the galaxy. Consecutive pointings followed a concentric ring-shaped pattern, adjusted to the shape of the PPAK bundle (see Figure 1). The observations of this galaxy spanned a period of three years. The area covered by all the observed positions accounts approximately for 34 arcmin$^2$, making this galaxy the widest spectroscopic survey ever made on a single nearby galaxy. The spectroscopic mosaic contains 11094 individual spectra.

With such dimensions, this galaxy allows us to study the 2D metallicity structure of the disc, the second order properties of its abundance distribution, and—as a very important byproduct—a complete 2D picture of the underlying stellar populations of the galaxy. Note that the linear physical scale that a single PPAK fibre samples at the assumed distance of the galaxy is ~120 pc. This scale can be compared to the physical diameter of a well-known HII region in our Galaxy, that is, the Orion nebula ($D \sim 8$ pc), or to the extent of what is considered prototypes of extragalactic giant HII regions, such as 30 Doradus ($D \sim 200$ pc) or NGC 604 ($D \sim 460$ pc). The area sampled by an individual fibre in the mosaic would subtend a fraction of a typical giant HII region in NGC 628, but the same area would fully encompass a number of small and medium size HII regions of the galaxy (see Figure 2).

The IFS analysis of NGC 628 was taken as a case study in order to explore different spectra extraction and analysis methodologies, taking into account the signal-to-noise of the data, the 2D spatial coverage, the physical meaning of the derived results, and the final number of analysed spectra. The analysis performed on this object represents an example of the potential and extent of studies based on IFS on nearby galaxies. In the first paper of the series ([29], hereafter Paper I), we present a study of the line emission and stellar continuum of NGC 628 by means of pixel-resolved maps across the disc of the galaxy. This study includes a qualitative description of the 2D distribution of the physical properties inferred from the line intensity maps and a comparison of these properties with both the integrated spectrum of the galaxy and the spatially resolved spectra. In the second article ([30], hereafter Paper II), we present a detailed, spatially resolved spectroscopic abundance analysis, based on different spectral samples extracted from the area covered by the IFS observations of NGC 628, and we define a spectra selection methodology specially conceived for the study of the nebular emission in IFU-based spectroscopic observations. This allows us to derive the gas chemistry distribution across the surface of the galaxy with unprecedented detail. In the third paper of the series (Sánchez-Blázquez et al., submitted; hereafter Paper III), we present a stellar population analysis of the galaxy, after applying spectral inversion methods to derive 2-dimensional maps of star-formation histories and chemical enrichment.

In Paper I, spatially resolved maps of the emission line intensities and physical properties were derived for NGC 628. Contrary to previous attempts to perform a 2D wide-field analysis based on narrow-band (or Fabry-Perot) imaging, which only allowed a basic analysis of the physical parameters and/or required assumptions on the line ratios included within individual filters (e.g., H$\alpha$), the emission line maps presented in this paper were constructed from individual (deblended) emission lines at any discrete spatial location of the galaxy, where enough signal-to-noise was found. This fact allowed investigating the point-to-point variation of the physical properties over a considerable area on the galaxy. Extinction, ionization, and metallicity-sensitive indicator maps were derived from reddening corrected emission line maps. In general, they show that the ionized gas in these spiral galaxies exhibits a complex structure, morphologically associated with the star-forming regions located along the spiral arms. The (thermal) ionization is stronger along the spiral arms, associated with the HII regions, and more intense in the outer than in the inner ones. Indeed, the surface SFR is an order of magnitude stronger in the outer HII regions, at distance larger than ~100 arcsec (4.5 kpc), than in the inner ones. Considering that in these outer regions there is a lower mass density, the growing rate of stellar mass is considerably larger there than in the inner ones. Therefore, the growth of the galaxy is dominated by the inside-out process.

The spatially resolved distribution of the abundance shows a clear gradient of higher oxygen metallicity values from the inner part to the outer part of the galaxy, and along the spiral arms (see right-panel of Figure 2). However, in some instances, the value of the oxygen abundance (and other physical properties like extinction and the ionization parameter) varies within what would be considered a classical well-defined HII region (or HII complex), showing some level of structure. Indeed, the 2D character of the data allows us to study the small-scale variation of the spectra within a given emitting area. The values of the emission line ratios measured using different extraction apertures vary considerably as a function of the aperture size, and the scatter of the central value is larger than the statistical error in the measurements, reflecting that this might in fact be a physical effect. By constructing 2D maps of the oxygen abundance distributions, we found that the 2D metallicity structure of the galaxy varies depending on the metallicity calibrator employed in order to derive the oxygen abundance. Different calibrators find regions of enhanced log(O/H) at spatial positions which are not coincident among them. This implies that the use of different empirical calibrations does not only reflect in a linear scale offset but may introduce spurious inhomogeneities. This information is usually lost in a simple radial abundance gradient, and that might be relevant when constructing a chemical evolution model based on a particular abundance determination (see Figure 3).

The emission line maps presented in Paper I proved to be useful in describing the general 2D properties of the galaxy. More robust conclusions were presented in Paper II, where we analysed specific individual regions across the disc of the galaxy, either by taking individual spectra above as a certain S/N threshold, or by coadding spectra with the same physical properties and comparing the results in the 2D context. With the first method we were able to identify regions of interstellar *diffuse* emission (see left panel of Figure 3), while with the second we created a *classic* catalogue of HII regions from a purely geometrical principle, that is, by coadding fibres considered to belong to the same morphological region.

Some highlights of this study (which also apply to the rest of the PINGS galaxies analysed so far) are the following.

(1) Despite the large number of spectra contained in the original observed mosaic, the final number of fibres containing analysable spectra of enough signal-to-noise for a spectroscopic study of the ionized gas represents only a reduced percentage of the total number of fibres contained in the full IFS mosaic. For the particular case of NGC 628, less than 10% of the total area sampled by the IFU observations is considered of sufficient quality.

(2) Independently of the abundance calibrator used, the metallicity distribution of NGC 628 is consistent with a nearly flat distribution in the innermost regions of the galaxy ($\rho/\rho_{25} < 0.2$), a steep negative gradient for $0.2 \lesssim \rho/\rho_{25} < 1$, and a shallow or nearly constant distribution beyond the optical edge of the galaxy, that is, implying a multimodality of the abundance gradient of NGC 628. The same feature is observed for the N/O versus $\rho$ distribution. The existence of this feature may be related to the differences in the 2D gas surface density and star-formation rate between the inner and outer disc which inhibits the formation of massive stars in the outer regions, causing a lack of chemical evolution in the outer disc compared with the inner regions.

(3) The observed dispersion in the metallicity at a given radius is neither a function of spatial position, nor due to low S/N of the spectra, and shows no systematic dependence on the ionization conditions of the gas, implying that the dispersion is real and is reflecting a true spatial physical variation of the oxygen content (see Figure 3).

(4) The values of the oxygen abundance derived from the integrated spectrum for each calibrator equal the abundance derived from the radial gradient at a radius $\rho \sim 0.4\rho_{25}$, confirming for this galaxy the previous results obtained for other objects, that is, that the integrated abundance of a normal disc

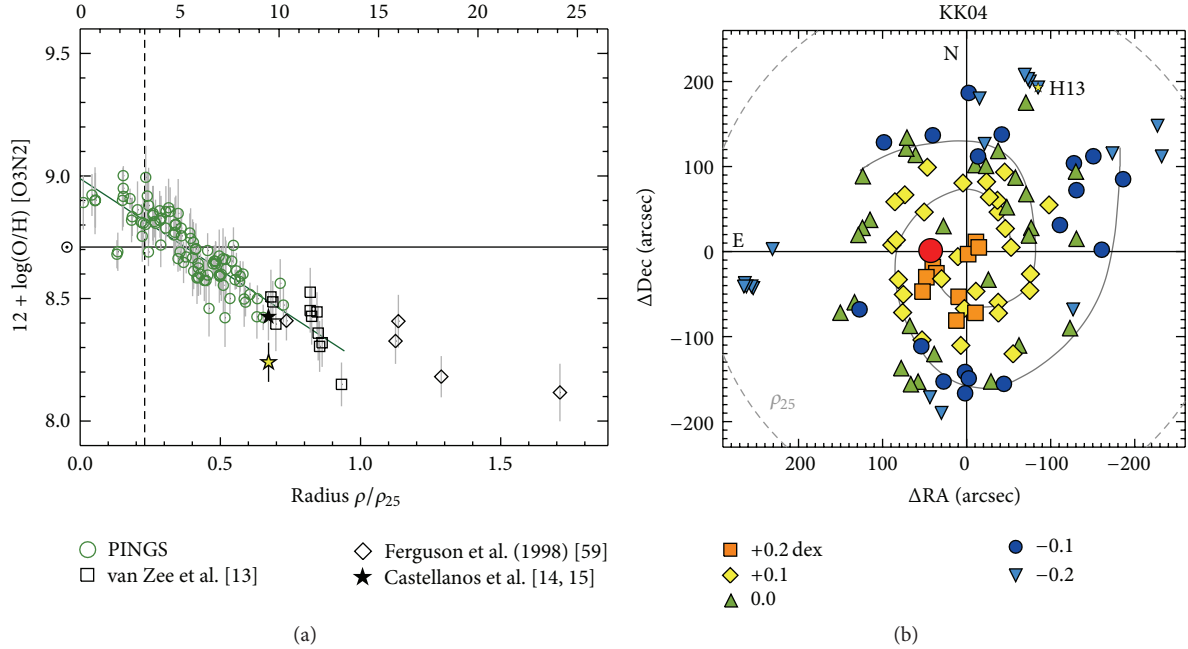

Figure 3: (a) Radial abundance gradient derived for NGC 628 based on the PINGS HII region catalogue (green symbols), and HII regions from the literature (black symbols) using the O3N2 calibrator. The horizontal grey lines correspond to the abundance derived using the integrated spectrum as reported in Paper I. The top $X$-axis values correspond to the projected radii in arcsec for the radial average data. Note the flattening of the gradient for innermost regions of the galaxy and for radii > $\rho_{25}$, that is, a multimodality of the abundance gradient. (b) 2D distribution of the oxygen abundance derived from the IFS H II regions catalogue of NGC 628 (plus selected HII regions from the literature), for the KK04 (top-left) metallicity calibrators. The shape and colours of the symbols correspond to the difference $\Delta\,[12 + \log(\mathrm{O/H})] \equiv \Delta\log(\mathrm{O/H})$ between the abundance obtained on each HII region with respect to the characteristic abundance $12 + \log(\mathrm{O/H})_{\rho=0.4\rho_{25}}$ of the same calibrator, grouped into bins of 0.0, ±0.1, 0.2 dex (e.g., +0.1 dex = 0.05 ≤ $\Delta\log(\mathrm{O/H})$ < 0.15). The large symbol in red colour stands for the location of the HII region with the maximum amount of $12 + \log(\mathrm{O/H})$ measured for that calibrator. The grey thick lines define the operational spiral arms of the galaxy. The dotted circle corresponds to the size of the optical radius $\rho_{25}$. Figure adapted from Rosales-Ortega et al. [30].

galaxy correlates with the characteristic gas-phase abundance measured at $\rho \sim 0.4\rho_{25}$.

(5) While trying to find axisymmetric variations of the metallicity content in the galaxy, we found slight variations between the central oxygen abundance and slopes for both the geometrical (quadrants) and morphological (arms) regions of the galaxy. However, these small variations fall within the expected errors involved in strong-line empirical calibrations (see Figure 4). If the radial trends in the ionization parameter and metallicity abundance were somewhat distinct, this would indicate that, to a certain extent, the physical conditions and the star-formation history of different-symmetric regions of the galaxy would have evolved in a different manner. Likewise, [32] found no evidence for significant large-scale azimuthal variations of the oxygen abundance across the whole disk of M 101 and marginal evidence for the existence of moderate deviations from chemical abundance homogeneity in the interstellar medium of this galaxy.

In the case of the stellar populations, in Paper III we derive maps of the mean (luminosity and mass weighted) age and metallicity that reveal a negative age gradient and the presence of structures such as a nuclear ring, previously seen in molecular gas (see Figure 5). The disc is dominated in mass by an old stellar component at all radii sampled by the IFS data, while the percentage of young stars increases with radius, as predicted in an inside-out formation scenario, where outer parts of the disc formed later due to the increasing timescales for gas infall with radius. We also detect an inversion of the metallicity gradient at the very centre of the galaxy (~1 kpc), where apparently there exists a ring of old stars at this distance, with a trend to younger ones at the very center. Similar results are found in the Milky Way (MW) using Open Clusters and Cepheids, that is, a clear bimodal gradient for the older population, with a flat outer plateau, and a more continuous gradient for the younger population (e.g., [33–36]). This behaviour has also been reported in other galaxies, mostly Sa/S0, where the inner regions of their bulges present bluer colors, consistent with younger stellar populations (e.g., [37]).

The relevance of this study regarding the nebular emission is that the young component shows a metallicity gradient that is very similar to that of the gas, and that is flatter than that of the old stars. Although the metallicity gradients for the young stars and the gas also show a break, this is much less prominent than for the old stars. The position of the break is more coincident with the corotation radius of the oval distortion than that of the spiral pattern, which is beyond the radius sampled by our data. We speculate about the possible origen of this break, the possibilities being due to star-formation variation with the spiral pattern speed or that is due to radial mixing produced by either the spiral arms, the oval distortion, or a coupling of both. We argue that NGC 628 could represent a good example of secular evolution due to the presence of a dissolving bar. In this scenario, the strong bar has funneled large amounts of gas into the central regions while radial flows induced in the disc have flattened the O/H gradient. Nuclear starbursts resulting from the gas sinking into the center contributed to the bulge's growth until enough mass was accreted to dissolve the bar by dynamical instabilities. The oval distortion observed in the central region could be the remains of the bar. Forthcoming studies analysing a sample of galaxies with different masses and showing different morphological features (e.g., bars of different strength, spiral arms with different morphologies, etc.) using, for example, the CALIFA survey that will help to elucidate the importance of the different mechanisms producing radial mixing in the galaxy discs.

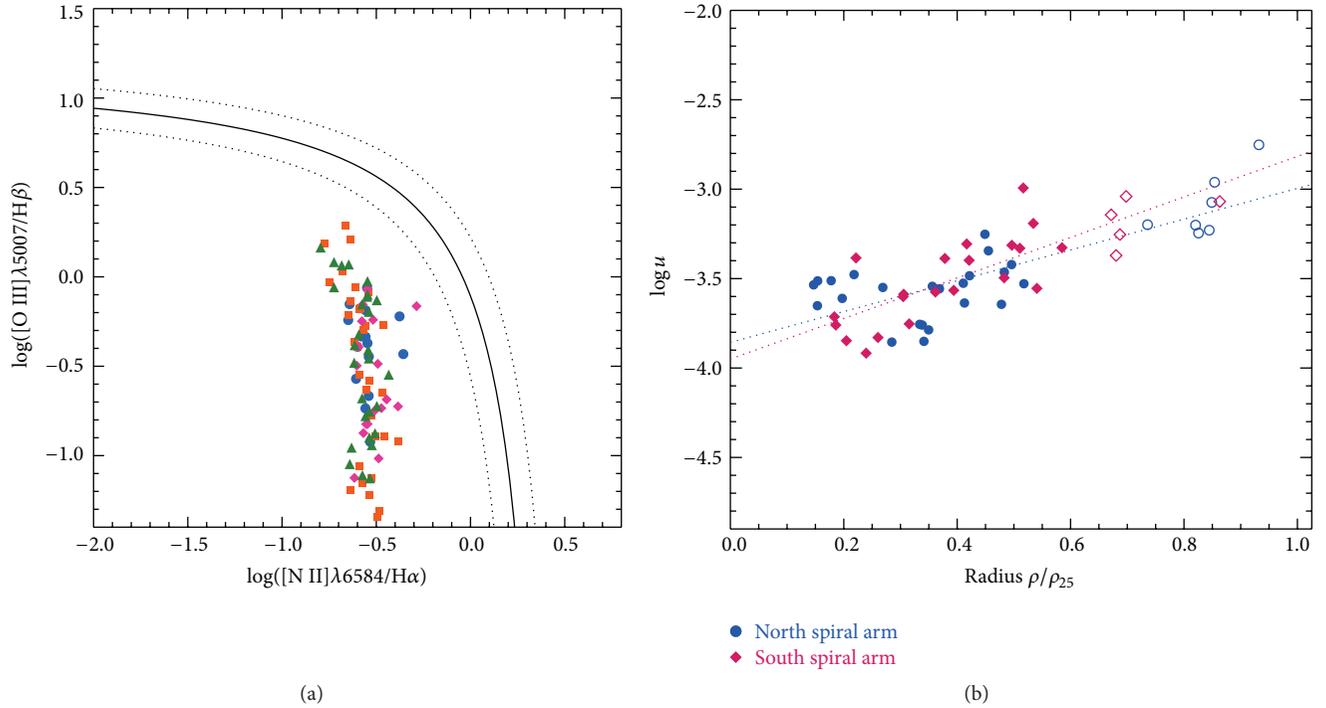

FIGURE 4: (a) BPT diagnostic diagram for HII regions coded according to the geometric position (quadrants) with respect to the an arbitrary axis drawn across the galaxy surface. The locus of different sectors does not show a clear trend or do not populate a clearly visible region on any diagram, compared to the rest of the quadrants. Points from all the regions are equally distributed within the cloud of points on each diagnostic diagram, indicating that the emission line ratios of the HII regions are not a function of azimuthal angle across the disc. (b) Radial gradients of the ionization parameter $\log u$ for morphologically selected HII regions of NGC 628. The panel shows the $\log u$ versus $\rho$ relation for the regions belonging to the north and south spiral arms of the galaxy. The difference between the two spiral arms resides in the slope of the gradient of $\log u$, for the north arm, and the values of $\log u$ increase moderately with galactocentric distance, while, for the south arm, the ionization parameter increases with a steeper slope, although within the errors of the linear fittings. Figure adapted from Rosales-Ortega et al. [30].

## 4. Hints of a Universal Abundance Gradient

IFS offers the possibility to analyse and study a single object in great detail, such as the case of NGC 628 described above. However, it also offers the unique chance of studying the spectroscopic properties of thousands of HII regions in a homogeneous way. We used our catalogue of HII regions introduced in Section 2 to characterize the radial trends and the physical properties of the HII regions of the galaxy sample. However, contrary to the case of NGC 628 where the HII regions on the disc of the galaxy were basically selected and extracted by-hand, the HII regions in these galaxies were detected, spatially segregated, and spectrally extracted using HIIexplorer [39], a new automatic procedure to detect HII regions, based on the contrast of the H$\alpha$ intensity maps extracted from the data cubes. Once detected, the algorithm provides with the integrated spectra of each individual segmented region. This change of paradigm is totally necessary when working with thousands of HII regions, contrary to the case of a handful of targets in classic long-slit spectroscopy. We detected a total of 2573 HII regions with good spectroscopic quality. This is by far the *largest* spatially resolved, nearby spectroscopic HII region survey ever accomplished. The emission lines were decoupled from the underlying stellar population using FIT3D [40], following a robust and well-tested methodology [20, 29]. Extinction-corrected, flux intensities of the stronger emission lines were obtained and used to select only star-forming regions based on typical BPT diagnostic diagrams. The final sample comprises 1896 high-quality, spatially resolved HII regions/aggregations of disc galaxies in the local Universe [39].

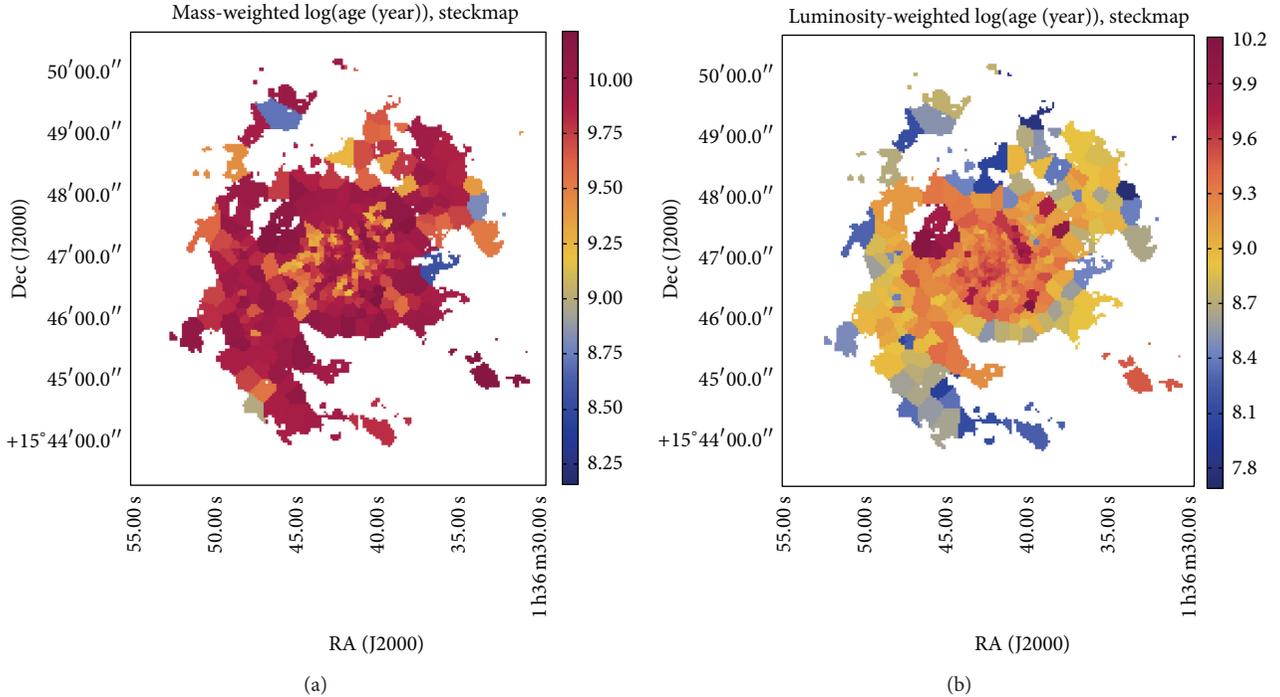

Figure 5: Mean age 2-D maps weighted by the mass (a) and by the light (b) of the stars. The different regions correspond to a Voronoi-tessellation binning scheme performed to the IFS mosaic of NGC 628. Figure adapted from Sánchez-Blázquez et al. (submitted).

It is well known that different spectroscopic properties of HII regions show strong variations across the area of disc galaxies. In particular, some of these parameters (e.g., oxygen abundance, EW[H$\alpha$], etc.), show a strong radial gradient, that in average indicates that more evolved, metal rich, stellar populations are located in the center of galaxies, and less evolved, metal poor ones are in the outer ones. Despite the several different studies describing these observational events, there is a large degree of discrepancy between the actual derived parameters describing the gradients: (i) slope of the gradient, (ii) average value and dispersion of the zero-point, and (iii) scale length of the truncation. In general, this is mostly due to different observational biases and the lack of a proper statistical number of analysed HII regions per galaxy.

For each galaxy of our sample we derived the correlation coefficient, the slope, and the zero point of a linear regression for a number of parameters showing radial distributions across the discs of the galaxies. For those properties showing a strong correlation, we investigated if the gradient was universal within our range of explored parameters. We found that, for the equivalent width of H$\alpha$ and the oxygen abundance, the slopes of the gradients are consistent with a Gaussian distribution; that is, the dispersion of values found for each individual galaxy is compatible with the average one, not showing strong statistical deviations. This implies that we can define a characteristic value for the slope and that we do not find a population of galaxies with slopes inconsistent with this normal distribution. The right panel of Figure 6 shows the radial density distribution for the oxygen abundance derived using the O3N2 indicator [28], once scaled to the average value at the effective radius for each galaxy. The radial distance was normalised to the effective radius of each galaxy. The solid line shows the average linear regression found for each individual galaxy. The red-dashed line shows the actual regression found for all the HII regions detected for all the galaxies.

Our results seem to indicate that there is a *universal* radial gradient for oxygen abundance and the equivalent width of H$\alpha$ when normalized with the *effective radii* of the galaxies; that is, they present a radial gradient that, statistically, has the same slope for all the galaxies in our sample. The derived slopes for each galaxy are compatible with a Gaussian random distribution and are independent of the morphology of the analysed galaxies (barred/nonbarred, grand-design/flocculent). This is one of the most important results in the abundance gradients of spiral galaxies, obtained thanks to the use of IFS.

## 5. The Local Mass-Metallicity Relation

The existence of a strong correlation between stellar mass and gas-phase metallicity in galaxies is a well-known fact. The mass-metallicity ($\mathcal{M}$-Z) relation is consistent with more massive galaxies being more metal-enriched; after the seminal work on this relationship by Lequeux et al. [41], it was firmly established observationally by Tremonti et al. ([42], hereafter T04) using the SDSS. However, there has been no major effort to test the $\mathcal{M}$-Z relation using *spatially resolved* information. We used our IFS observations in order to test the distribution of mass and metals *within* the discs of the galaxies. We derived the (luminosity) surface mass density ($\Sigma_{\text{Lum}}$, $M_\odot\,\text{pc}^{-2}$) within the area encompassed by our IFS-segmented HII regions, using the prescriptions given by Bell and de Jong [43] to convert $B$-$V$ colors into a $B$-band mass-to-light ratio ($M/L$).

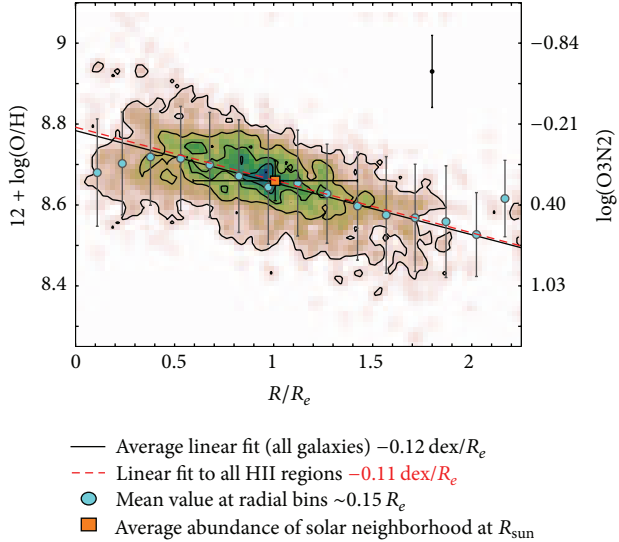

Figure 6: Radial oxygen abundance density distribution for the whole HII region spectroscopic sample discussed in the text. The first contour indicates the mean density, with a regular spacing of four times this value for each consecutive contour. The light-blue solid circles indicate the mean value (plus $1 - \sigma$ errors) for each consecutive radial bin of $\sim 0.15\,R_e$. The average error of the derived oxygen abundance is shown by a single error bar located at the top-right side of the panel. The solid-orange square indicates the average abundance of the solar neighbourhood, at the distance of the Sun to the Milky-Way galactic center. The lines represent linear fits to all galaxies (black) and all HII regions (dotted red) independently, showing a *universal* slope $\sim -0.1\,\text{dex}/R_e$. Figure is adapted from Rosales-Ortega et al. [38].

The left panel of Figure 7 shows the striking correlation between the local surface mass density and gas metallicity for our sample of nearby HII regions, that is, the *local $\mathcal{M}$-Z relation*, extending over $\sim 3$ orders of magnitude in $\Sigma_{\text{Lum}}$ and a factor $\sim 8$ in metallicity [38]. The notable similarity with the global $\mathcal{M}$-Z relation can be visually recognised with the aid of the blue lines which stand for the [42] fit ($\pm 0.2$ dex) to the global $\mathcal{M}$-Z relation, shifted arbitrarily both in mass and metallicity to coincide with the peak of the HII region $\mathcal{M}$-Z distribution. Other abundance calibrations were tested obtaining the same shape (and similar fit) of the relation.

In addition, we find the existence of a more general relation between mass surface density, metallicity, and the equivalent width of H$\alpha$, defined as the emission-line luminosity normalized to the adjacent continuum flux, that is, a measure of the SFR per unit luminosity [44]. This functional relation is evident in a 3D space with orthogonal coordinate axes defined by these parameters, consistent with |EW(H$\alpha$)| being

inversely proportional to both $\Sigma_{\text{Lum}}$ and metallicity, as shown in Figure 8. As discussed in Rosales-Ortega et al. [38], we interpret the local $\mathcal{M}$-Z-EW(H$\alpha$) relation as the combination of (i) the well-known relationships between both the mass and metallicity with respect to the differential distributions of these parameters found in typical disc galaxies, that is, the *inside-out* growth, and (ii) the fact that more massive regions form stars faster (i.e., at higher SFRs), thus earlier in cosmological times.

In order to test whether the global $\mathcal{M}$-Z relation observed by [42] using SDSS data is a reflection (aperture effect) of the local HII region mass-density versus metallicity relation, we perform the following exercise. We simulate a galaxy with typical $M_B$ and $B$-$V$ values drawn from flat distributions in magnitude ($-15$ to $-23$) and colour ($\sim 0.4$–1). A redshift is assumed for the mock galaxy, drawn from a Gaussian distribution with mean $\sim 0.1$ and $\sigma = 0.05$, with a redshift cut $0.02 < z < 0.3$ in order to resemble the SDSS [42] distribution. The mass of the galaxy is derived using the integrated $B$-band magnitudes, $B$-$V$ colours, and the average $M/L$ ratio following Bell and de Jong [43]. The metallicity of the mock galaxy is derived using the local $\mathcal{M}$-Z relation within an aperture equal to the SDSS fiber (3 arcsec), that is, the metallicity that corresponds to the mass density surface at this radius. The process is repeated over 10,000 times in order to obtain a reliable distribution in the mass and metallicity of the mock galaxies.

The right panel of Figure 8 shows the result of the simulation, that is, the distribution of the mock galaxies in the $\mathcal{M}$-Z parameter space. We reproduce—with a *remarkable* agreement—the overall shape of the global $\mathcal{M}$-Z relation assuming a local $\mathcal{M}$-Z relation and considering the aperture effect of the SDSS fiber. The overlaid lines correspond to the [42] fit (black) and the Kewley and Ellison [45] $\pm 0.2$ dex relation (blue), for which the agreement is extremely good over a wide range of masses. The result is remarkable considering that we are able to reproduce the global $\mathcal{M}$-Z relation over a huge dynamical range, using a local $\mathcal{M}$-Z relation derived from a galaxy sample with a restricted range in mass ($9.2 < \log M_{\text{Lum}} < 11.2$) and metallicity ($8.3 < 12 + \log(\text{O/H}) < 8.9$), indicated by the rectangle shown in the right panel of Figure 8.

Therefore, by using the power of IFS applied to a sample of nearby galaxies we demonstrate the existence of a *local* relation between the surface mass density, gas-phase oxygen abundance, and |EW(H$\alpha$)| in $\sim 2000$ spatially resolved HII regions of the Local Universe. The projection of this distribution in the metallicity versus $\Sigma_{\text{Lum}}$ plane—the *local $\mathcal{M}$-Z relation*—shows a tight correlation expanding over a wide range in this parameter space. We use the local $\mathcal{M}$-Z relation to reproduce the global $\mathcal{M}$-Z relation by means of a simple simulation which considers the aperture effects of the SDSS fiber at different redshifts.

Note that the "local" $\mathcal{M}$-Z-|EW(H$\alpha$)| relation is conceptually different from the "global" $\mathcal{M}$-Z-SFR relation proposed by Lara-López et al. ([46], dubbed FP), Mannucci et al. ([47], dubbed FMR), or Hunt et al. [48], based on the integrated spectra of galaxies (the basic difference between these relations is the proposed *shape* in the 3D distribution, that is, a

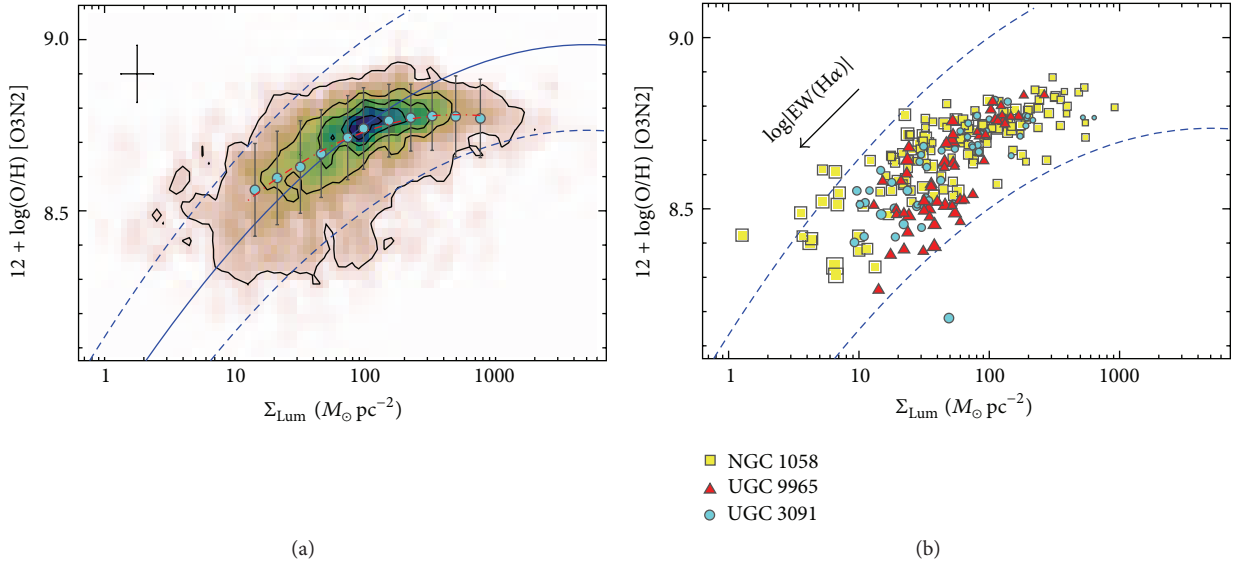

FIGURE 7: (a) The relation between surface mass density and gas-phase oxygen metallicity for ~2000 HII regions in nearby galaxies, the *local* $\mathcal{M}$-Z relation. The first contour stands for the mean density value, with a regular spacing of four time this value for each consecutive contour. The blue circles represent the mean (plus 1$\sigma$ error bars) in bins of 0.15 dex. The red dashed-dotted line is a polynomial fit to the data. The blue lines correspond to the [42] relation (±0.2 dex) scaled to the relevant units. Typical errors for $\Sigma_{\mathrm{Lum}}$ and metallicity are represented. (b) Distribution of HII regions along the local $\mathcal{M}$-Z relation for three galaxies of the sample at different redshifts. The size of the symbols is linked to the value of |EW(H$\alpha$)|, being inversely proportional to $\Sigma_{\mathrm{Lum}}$ and metallicity as shown. Figure is adapted from Rosales-Ortega et al. [38].

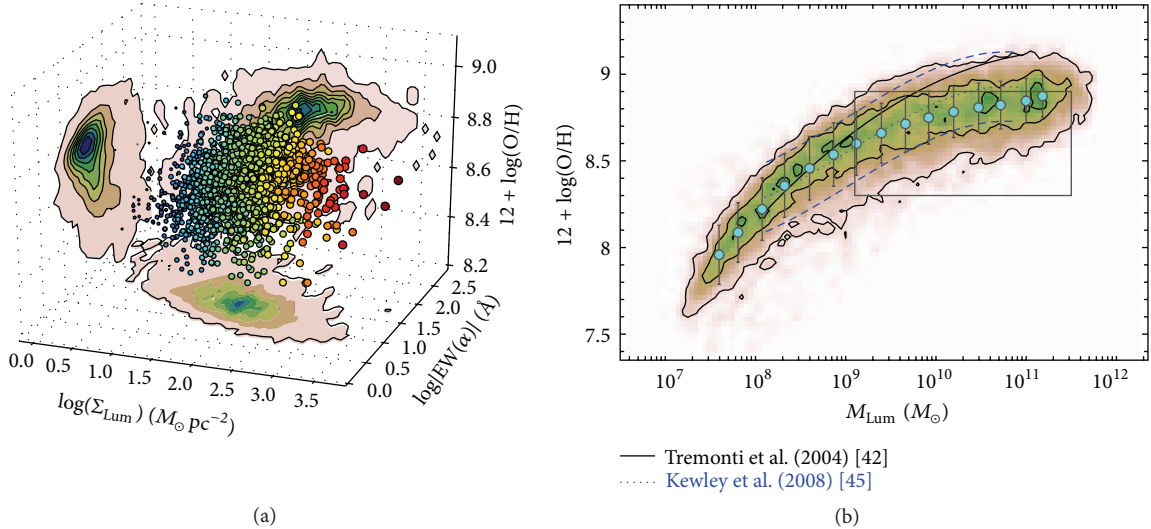

FIGURE 8: (a) 3D representation of the local $\mathcal{M}$-Z-EW(H$\alpha$) relation. The size and color scaling of the data points are linked to the value of log$\Sigma_{\mathrm{Lum}}$ (i.e. low-blue to high-red values). The projection of the data over any pair of axes reduces to the local $\mathcal{M}$-Z, $\mathcal{M}$-EW(H$\alpha$), and metallicity-EW(H$\alpha$) relations. An online 3D animated version is available at http://tinyurl.com/local-mass-metallicity. (b) Distribution of simulated galaxies in the $\mathcal{M}$-Z plane assuming a *local* $\mathcal{M}$-Z relation and considering the aperture effect of the SDSS fiber, as explained in the text. The contours correspond to the density of points, while the circles represent the mean value (plus 1$\sigma$ error bars) in bins of 0.15 dex. The black line stands for the [42] fitting, while the blue lines correspond to the Kewley and Ellison [45] ±0.2 dex relation. The rectangle encompasses the range in mass and metallicity of the galaxy sample. Figure is adapted from Rosales-Ortega et al. [38].

surface or a plane). However, the obvious parallelism between these two scaling relations deserves a discussion. While the "local" $\mathcal{M}$-Z-|EW(H$\alpha$)| relation is related to the intrinsic physics involved in the growth of the galaxy disc in an *inside-out* scenario, the existence of the "global" $\mathcal{M}$-Z-SFR relation is explained, according to Mannucci et al. [47], by the interplay of infall of pristine gas and outflow of enriched material at different redshifts epochs, supporting the smooth accretion scenario, where galaxy growth is dominated by continuous accretion of cold gas in the local Universe. However, Sánchez et al. [49], using CALIFA data, found no secondary relation of the mass and metallicity with the SFR other than the one induced by the primary relation of this quantity with the stellar mass. The same was found with respect to the specific SFR rate. The results by Sánchez et al. [49] agree with a scenario in which gas recycling in galaxies, both locally and globally, is much faster than other typical timescales, such like that of gas accretion by inflow and/or metal loss due to outflows. In essence, late-type/disc-dominated galaxies seem to be in a quasi-steady situation, with a behavior similar to the one expected from an instantaneous recycling/closed-box model.

In this scenario, the inner regions of the galaxy form first and faster, increasing the gas metallicity of the surrounding interstellar medium. As the galaxy evolves and grows with time, the star-formation progresses radially creating a radial metallicity gradients in the disk of spirals. Mass is progressively accumulated at the inner regions of the galaxy, raising the surface mass density and creating a bulge, with corresponding high metallicity values but low SSFR (low $|EW(H\alpha)|$), that is, an "inside-out" galaxy disk growth. In such a case, the local $\mathcal{M}$-Z relation would reflect a more fundamental relation between mass, metallicity, and star-formation efficiency as a function of radius, equivalent to a local *downsizing* effect, similar to the one observed in individual galaxies. Following this reasoning, the origin of the global $\mathcal{M}$-Z relation can be explained as the combined effect of the existence of the local $\mathcal{M}$-Z relation, an aperture bias due to the different fibers covering factors of the spectroscopic surveys from which the FMR and FP were derived (as second-order effect), and a possible selection of a bias of the galaxy populations which are most common at a particular redshift, and may not reflect the physics of how galaxies evolve. Supporting evidence in favour of the *inside-out* scenario of galaxy growth comes from the recent analysis of the spatially resolved history of the stellar mass assembly in galaxies of the local Universe [50]. In summary, the existence of the M-Z-SFR relation could also be interpreted as a scaled-up version of the local M-Z-sSFR relation in the distribution of star-forming regions across the discs of galaxies as described in Rosales-Ortega et al. [38] and confirmed by Sánchez et al. [49]; that is, the relationship is not primary, but obtained from the sum of a number of local linear relations (and their deviations) with respect to the galaxy radius.

## 6. Conclusions

The emergence of a new generation of instrumentation, that is, multiobject and integral field spectrometers with large fields of view, capable of performing emission-line surveys based on samples of hundreds of spectra in a 2D context, are revolutionising the methods and techniques used to study the gas-phase component of star-forming galaxies in the nearby Universe (objects which were typically studied with small samples based on long-slit spectroscopy). A new body of results is coming out from these studies, opening up a new frontier of studying the 2D structure and intrinsic dispersion of the physical and chemical properties of the discs of nearby spiral galaxies. In this paper we review some of the projects that in the last years tackled for the first time the problem of obtaining spatially resolved spectroscopic information of the gas in nearby galaxies. PINGS represented the first endeavour to obtain full 2D coverage of the discs of a sample of spiral galaxies in the nearby Universe. The PINGS sample covered different galaxy types, including normal, lopsided, interacting, and barred spirals with a good range of galactic properties and star-forming environments, with multiwavelength public data. The spectroscopic data set comprises more than 50 000 individual spectra, covering an observed area of nearly 80 arcmin$^2$, an observed surface without precedents by an IFS study by the time.

The IFS analysis of NGC 628, the largest spectroscopic mosaic on a single galaxy, was taken as an example of the new methodology and analysis that could be performed with a large spectroscopic database for a single object. The contribution of PINGS also resides in defining a self-consistent methodology in terms of observation, data reduction and analysis for present and future IFS surveys of the kind, as well as a whole new set of visualization and analysis software that has been made public to the community (e.g., [51, 52]). Despite all the complexities involved in the observations, data reduction, and analysis, PINGS proved to be feasible. In less than a three-year period, it was possible to build a comprehensive sample of galaxies with a good range of galactic properties and available multiwavelength ancillary data, maximising both the original science goals of the project and the possible archival value of the survey. In fact, the science case of the PINGS project was the inspiration for the ongoing CALIFA survey. The face-on spirals from Mármol-Queraltó et al. [21] were part of the feasibility studies for the CALIFA survey. On completion, CALIFA will be the largest and the most comprehensive wide-field IFU survey of galaxies carried out to date. It will thus provide an invaluable bridge between large single aperture surveys and more detailed studies of individual galaxies.

*6.1. Results from Other IFU Projects on Star-Forming Galaxies.* Other projects have followed this initiative; for example, the Mitchell spectrograph instrument at McDonald Observatory (a.k.a VIRUS-P) is currently used to carry out two small IFS surveys, namely, VENGA [53] and VIXENS [54]. VENGA (VIRUS-P Exploration of Nearby Galaxies) is an integral field spectroscopic survey, which maps the discs of 30 nearby spiral galaxies, in a very similar manner to PINGS in terms of spectral coverage, resolution, and area sampled (3600 Å–6800 Å, ~5 Å FWHM, ~$0.7R_{25}$) although with a different spatial resolution (5.6 arcsec FWHM). Their targets span a wide range in Hubble type, star-formation activity, morphology, and inclination. Likewise PINGS, the VENGA group used the data cubes of their observations to produce 2D maps of the star-formation rate, dust extinction, electron density, stellar population parameters, the kinematics and chemical abundance of both stars and ionized gas, and other physical quantities derived from the fitting of the stellar spectrum and the measurement of nebular emission lines. Their first results focus on (1) the spatially resolved star-formation law of NGC 5194 where they give support to the evidence for a low, and close to constant, star-formation efficiency (SFE = $\tau^{-1}$) in the molecular component of the interstellar medium [55] and (2) using IFS observations of NGC 628, they measure the radial profile of the $^{12}$CO(1–0) to $H_2$ conversion factor ($X_{CO}$) in this galaxy and study how changes in $X_{CO}$ follow changes in metallicity, gas density, and ionization parameter [56], and also they use the IFS data to propose a new method to measure the inclination of nearly face-on systems based on the matching of the stellar and gas rotation curves using asymmetric drift corrections [53]. In the case of VIXENS (VIRUS-P Investigation of the eXtreme ENvironments of Starburst), their goal of our survey is to investigate the relation between star-formation and gas content in the extreme environments of interacting galaxy pairs and mergers on spatially resolved scales of 0.2–0.8 kpc, by using IFS of 15 interacting/starburst galaxies. VIXENS will make extensive use of multiwavelength data in order to investigate the star-formation in this object, including data from Spitzer, GALEX, IRAM, CARMA archival CO, and Hi maps. These projects and datasets are clearly focused on specific science questions, adopting correspondingly optimized sample selection criteria and also observing strategies.

Other surveys in the local Universe using the power of IFS for a detailed study of nearby galaxies include the next generation surveys like Sydney university AAO MOS IFU [57] (SAMI) and Mapping Nearby Galaxies at APO, PI: Kevin Bundy, IPMU (MaNGA), or the new generation instrumentation for Very Large Telescope (VLT, ESO) like Multi Unit Spectroscopic Explorer [58] (MUSE), which aim at studying the the chemical and dynamical evolution history and dark matter contents of galaxies, the physical role of environment in galaxy evolution, when, where, and why does star-formation occur, and so forth, based on spatially resolved spectroscopic surveys of $10^4$–$10^5$ galaxies. The continuous coverage spectra provided by the imaging spectroscopy technique employed in these projects are already allowing us to study the small and intermediate linear scale variation in line emission and the gas chemistry for a statistically representative number of galaxies of the nearby Universe. The primary motivation common to all of these observational efforts is to use this information to link the properties of high redshift galaxies with those we see around us today and thereby understand the physical processes at play in the formation and evolution of galaxies. The power and importance of all these projects resides in the fact that they will provide an observational anchor of the spatially resolved properties of the galaxies in the local Universe, which will have a potential impact in the interpretation of observed properties at high redshift from new generation facilities, such as the James Webb Space Telescope (JWST), the Giant Magellan Telescope (GMT), or the European Extremely Large Telescope (E-ELT), projects that will hopefully revolutionise the understanding of our Universe in future years.

## Acknowledgments

Based on observations collected at the Centro Astronómico Hispano-Alemán (CAHA) at Calar Alto, operated jointly by the Max-Planck Institut für Astronomie and the Instituto de Astrofísica de Andalucía (CSIC) Fernando Fabián Rosales-Ortega acknowledges the Mexican National Council for Science and Technology (CONACYT) for financial support under the Programme Estancias Posdoctorales y Sabáticas al Extranjero para la Consolidación de Grupos de Investigación, 2010–2012. The author also acknowledges financial support for the ESTALLIDOS collaboration by the Spanish Ministerio de Ciencia e Innovación under Grant AYA2010-21887-C04-03.

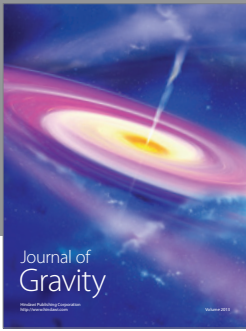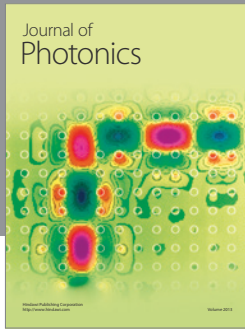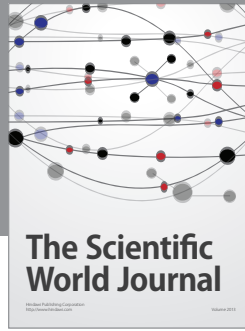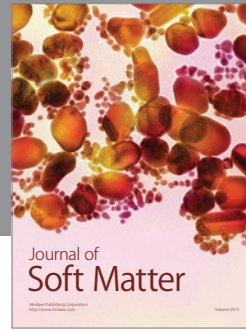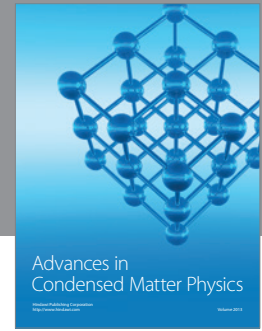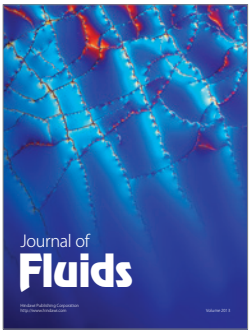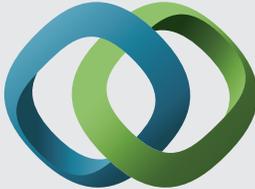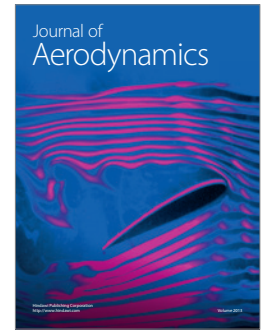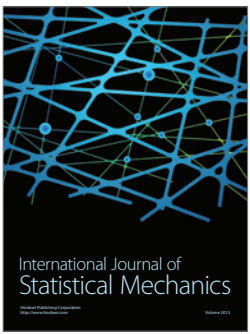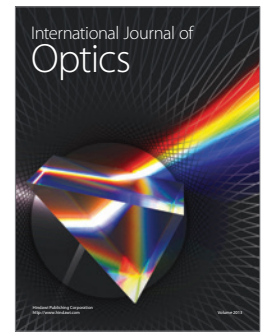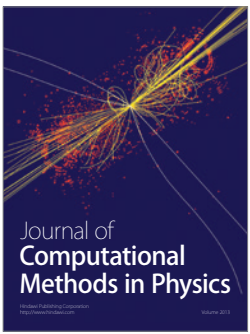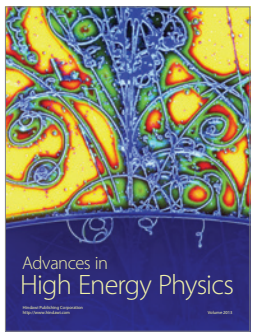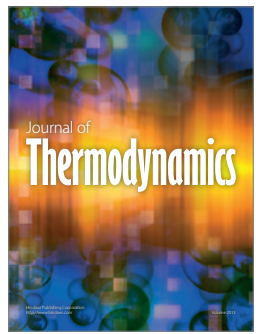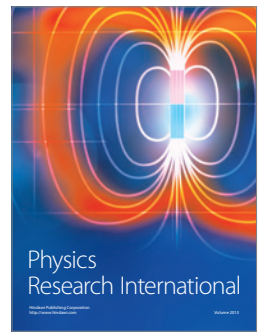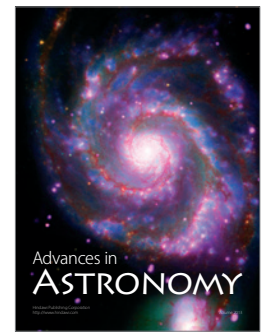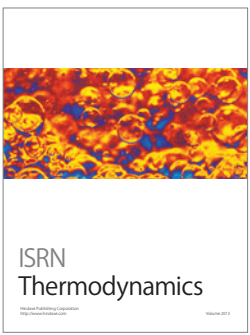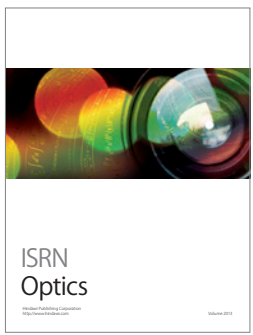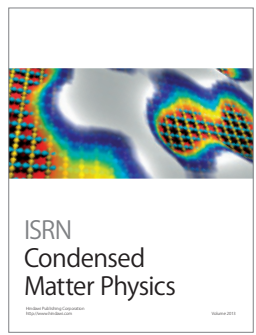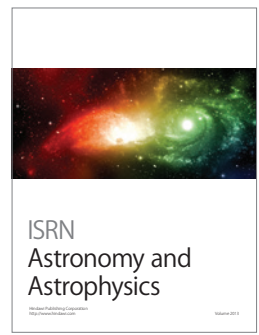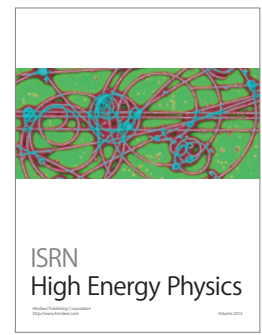